\documentclass[amsmath,amssymb,twocolumn,prl,superscriptaddress]{revtex4}

\usepackage{graphicx}
\usepackage{acronym}

\newcommand{\beq}{\begin{equation}}
\newcommand{\eeq}{\end{equation}}
\newcommand{\beqa}{\begin{eqnarray}}
\newcommand{\eeqa}{\end{eqnarray}}

\def\tmc{T_\mathrm{MC}}
\def\xidyn{\xi_\mathrm{dyn}}
\def\ximos{\xi_\mathrm{mos}}

\begin{document}

\title{Mosaic multi-state scenario vs.\ one-state description of
  supercooled liquids}

\author{Andrea Cavagna} \affiliation{Centre for Statistical Mechanics
and Complexity, ISC, INFM-CNR, Via dei Taurini 19, 00185 Roma, Italy}

\author{Tom\'as S. Grigera} \affiliation{Instituto de Investigaciones
  Fisicoqu\'\i{}micas Te\'oricas y Aplicadas (INIFTA) and Departamento
  de F\'\i{}sica, Facultad de Ciencias Exactas, Universidad Nacional
  de La Plata, c.c. 16, suc. 4, 1900 La Plata, Argentina.}
  \affiliation{Consejo Nacional de Investigaciones Cient\'\i{}ficas y
  T\'ecnicas, Argentina}

\author{Paolo Verrocchio}
\affiliation{Dipartimento di Fisica, Universit\`a di Trento, via
Sommarive 14, 38050 Povo, Trento, Italy.}  
\affiliation{SOFT, INFM-CNR, Universit\`a di Roma ``La Sapienza'', 
00185, Roma, Italy.}
\affiliation{Instituto de Biocomputaci\'on y F\'{\i}sica de Sistemas
Complejos (BIFI), Zaragoza, Spain.}

\date{December 19, 2006}

\begin{abstract}
  According to the mosaic scenario, relaxation in supercooled liquids
  is ruled by two competing mechanisms: surface tension, opposing the
  creation of local excitations, and entropy, providing the drive to
  the configurational rearrangement of a given region. We test this
  scenario through numerical simulations well below the Mode Coupling
  temperature. For an equilibrated configuration, we freeze all the
  particles outside a sphere and study the thermodynamics of this
  sphere. The frozen environment acts as a pinning field. Measuring
  the overlap between the unpinned and pinned equilibrium
  configurations of the sphere, we can see whether it has switched to
  a different state. We do not find any clear evidence of the mosaic
  scenario. Rather, our results seem compatible with the existence of
  a single (liquid) state. However, we find evidence of a growing
  static correlation length, apparently unrelated to the mosaic one.
\end{abstract}

\maketitle

It is common opinion that in finite dimension a divergence of a
relaxation time $\tau$ at nonzero temperature is associated to a
diverging characteristic length $\xi$. The idea is that when this
length increases, relaxation proceeds through the rearrangement of
ever larger regions, taking a longer and longer time.  The relation
between $\tau$ and $\xi$ depends on the physical mechanism of
relaxation. Two main mechanisms are activated relaxation of a
$\psi$-dimensional droplet of size $\xi$, giving $\tau \sim \exp(A
\xi^\psi/T)$, and critical slowing down, where $\tau \sim \xi^z$
\cite{ZINNJUSTIN}.

Glass-forming liquids are tricky: relaxation times grow spectacularly
(more than ten decades) upon lowering the temperature, without clear
evidence of a growing {\em static} cooperative lenght.  In particular,
density fluctuations are thought to remain correlated over short
distances close to the glass transition (although there are some
indication that energy fluctuations might develop larger
correlations~\cite{Fernandez06,Matharoo06}).  Thus the concept of
\emph{dynamic} heterogeneities is central to several theories of the
glass transition \cite{Parisi99,garrahan02,whitelam04,cugliandolo03},
where the role of order parameter is played by dynamic quantities such
as local time correlators, which become correlated over the growing
dynamic lenght scale $\xidyn$. No thermodynamic singularity is present
in these theories. Dynamic singularities are also typically absent at
finite temperatures, with the notable exception of \ac{MCT}
\cite{Goetze92}, recently recast in terms of dynamic
heterogeneities~\cite{biroli04}. Note, however, that the experimental
values of $\xidyn$ \cite{berthier05,EDIGER00,ISRAELOFF00,sillescu99}
are barely in the nm range, the same as density
correlations~\cite{Gaskell96}.

The \ac{MS}~\cite{kirkpatrick89,Xia01,Lubchenko06}, working within the
conceptual framework of nucleation theory, identifies on the other
hand a \emph{static} correlation length.  Deeply rooted in the physics
of mean-field spin glasses, the \ac{MS} crucially assumes the
existence of exponentially many inequivalent states $\exp(N\Sigma)$,
below the mode coupling temperature $\tmc$ ($\Sigma$ is called
complexity or configurational entropy, and $N$ is the size of the
system). Suppose the system is in a state $\alpha$ and ask: what is
the free energy cost for a region of linear size $R$ to rearrange into
a different state $\beta$ with the same free energy?  According to the
\ac{MS}, there are two opposing contributions: a surface cost due to
the mismatch of $\alpha$ and $\beta$, proportional to $\Upsilon
R^\theta$, where $\Upsilon$ is a generalized surface tension
($\theta\leq d-1$), and an entropic gain proportional to $T\Sigma
R^d$, arising from the fact that the larger the region, the higher the
number of possible rearrangements. This is similar to nucleation
theory, with $T\Sigma$ playing the role of the free energy difference
between the two phases. For small droplets, the surface contribution
dominates, whereas for large $R$ the volume contribution wins, and
eventually the rearrangement occurs.  Thus, similarly to nucleation
theory, a length scale emerges, $\ximos \equiv
(\Upsilon/T\Sigma)^\frac{1}{d-\theta}$ fixed by the balance of the two
contributions. A droplet with $R<\ximos$ may flip to a different
state, but the pinning field provided by the surrounding system will
make it flip back; for $R>\ximos$ the region is {\it
  thermodynamically} favoured to flip, but the larger $R$ the longer
the {\it time} the rearrangement takes. $\ximos$ thus acts as the
typical cooperative length. Furthermore, by noting that $\ximos$
diverges at the temperature $T_0$ where $\Sigma$ vanishes, the \ac{MS}
provides a rationale for the Vogel-Fulchner-Tamman (VFT) law for the
relaxation time, $\tau \sim\exp[\Delta/(T-T_0)]$.

In this Letter we report a numerical test of the \ac{MS}. For this
purpose, the formulation of ref.~\onlinecite{biroli04b} is
particularly convenient. Imagine picking a reference equilibrium
configuration and freezing all particles except those within a sphere
of radius $R$ (containing $M$ particles). This region is thus embedded
in a very large box of frozen particles which act as a pinning field.
The key point of the \ac{MS} is that a sufficiently large sphere will
be thermodynamically favored to flip to a different state. If the
reference configuration is in state $\alpha$, and calling $\beta$ one
of the exponentially many different states, the \ac{MS} gives
\cite{biroli04b}
\begin{equation}
  \label{jogabonito}
  p_{\alpha\alpha}(R)= \frac{1}{Z} \exp(\Upsilon R^\theta/T), \qquad
  p_{\alpha\beta}(R)= \frac{1}{Z} \exp(\Sigma R^d),
\end{equation}
as the probabilities for the sphere to remain in the reference state
$\alpha$ and to flip to $\beta$, respectively (the normalization is
$Z=\exp(\Upsilon R^\theta/T)+\exp(\Sigma R^d)$).  The mosaic
fragmentation of a state into regions of size $\ximos$ corresponds to
an exponentially sharp jump from $p_{\alpha\alpha}\sim 1$ to
$p_{\alpha\alpha}\sim 0$ at $R \sim \ximos$~\cite{biroli04b} .  Let us
call the configuration of the sphere thermalized with a given
environment the {\it pinned} equilibrium state of the sphere. If we
define a suitable overlap $q$ to measure the similarity in phase space
between this pinned state and the reference state $\alpha$, the
\ac{MS} predicts,
\begin{equation}
q_\mathrm{MS}(R) = p_{\alpha\alpha}(R) q_{\alpha\alpha} +
                  p_{\alpha\beta}(R) q_0,
\label{barnetta}
\end{equation} 
where $q_{\alpha\alpha}$ is the self-overlap of state $\alpha$ and
$q_0$ is the typical overlap between different states.  Thus within
the \ac{MS} one expects a first order transition~\cite{Superfranz06},
where $q_\mathrm{MS}(R)$ drops (or at least makes a crossover) at
$\ximos$ from $q_{\alpha\alpha}$ to $q_0$. Note finally that for large
$R$ the overlap decays exponentially, $q_\mathrm{MS}(R) - q_0 \sim
\exp[-(\Sigma R^d-YR^\theta/T)]$.

We have realized this {\sl gedanken} experiment for the soft-sphere
binary mixture \cite{Bernu87}, a simple fragile glassformer (see
ref.~\cite{jack05} for an attempt to find a mosaic length-scale in
spin models using this formulation). We used the swap Monte Carlo
algorithm \cite{grigera01} (for simulation details see
ref.~\onlinecite{Grigera04}). Equilibration time is considerably
shortened, so we can perform equilibrium simulations of large systems
well below $\tmc =$0.226~\cite{Bernu87,Yu04}. This is important, since
in the standard MS $\tmc$ acts like a spinodal
temperature~\cite{kirkpatrick89}, above which there are not many
states and the surface tension is zero \footnote{Note that according
  to~\cite{Montanari06} the mosaic mechanism is active even {\em
    above} $\tmc$}.  We first equilibrated systems both of $2048$ and
of $16384$ particles, then picked reference configurations used to
equilibrate $M$ particles within of the sphere while keeping the
remaining frozen. We studied $M= 5$, 20, 50, 100, 200, 400, 800, 1600,
3200, and 5500 at temperatures $T/\tmc=4.42$, 2.13, 1.54, 1.15, 0.94,
and 0.89. The largest $R$ was $10.95$, one order of magnitude larger
than the particle size. Though reminiscent of the work of
ref.~\onlinecite{Scheidler02}, we focus on static rather than dynamic
quantities.

To introduce the overlap between the reference state $\alpha$ and the
pinned state of a sphere of radius $R$, we divide the space in cells
of side $l$ and define,
\begin{equation}
q(R)=\frac{1}{M} \sum_i \langle n_i^{(\alpha)}\rangle\langle
n_i^{(\mathrm{pin})}\rangle,
\label{elizondo}
\end{equation}
where $n_i$ is the occupation number of cell $i$. The sum runs over
all the cells in the sphere and $M=4/3\pi R^3 \rho$. Occupation
numbers are averaged over many independent configurations, both of
state $\alpha$ (between $4$ and $16$) and the pinned state ($10$ to
$100$). The overlap of two identical configurations of the sphere is
$q=1$, whereas for two independent configurations $q=q_0=\epsilon$,
with $\epsilon=\rho l^3$.  $l$ is such that $\epsilon \ll 1$, but
larger than the typical vibrational amplitude of the particles
\cite{Grigera04}. Here $\epsilon = 0.06$.

\begin{figure}
\includegraphics[clip,width=.75\columnwidth,angle=270]{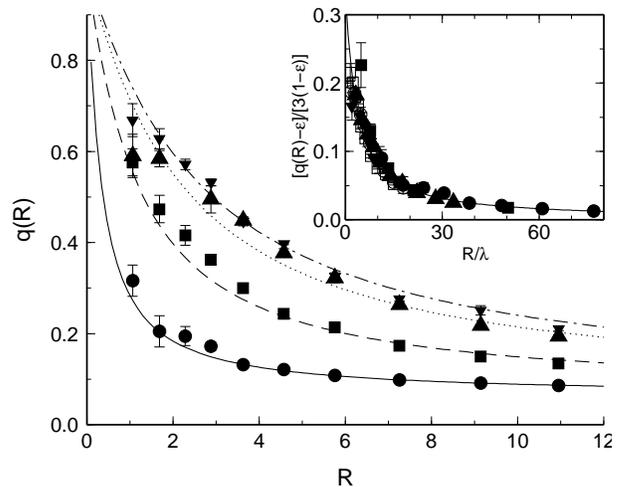}
\caption{Overlap $q(R)$ for $T/T_\mathrm{MCT}=$4.42 (circles),
  1.54 (squares), 0.94 (triangles), 0.89 (inverted triangles). The
  largest value of $R$ corresponds to $M=5500$ particles in the
  sphere. Lines are the fit from the 1S argument (Eq.~\ref{naive}).
  Inset: data for all temperatures scaled according to
  Eq.~(\ref{naive}). }
\label{qR}
\end{figure}

Our results for $q(R)$ are shown in Fig.~\ref{qR} at four different
temperatures, two above and two below $\tmc$.  The first feature we
notice is that $q(R)$ is always a smooth function, with no clear jump,
nor evident crossover values of $R$. Yet, even at the lowest
temperature, $q$ becomes as low as $0.2$ at the largest value of $R$,
showing that the sphere has in fact largely decorrelated with respect
to the reference state $\alpha$. This means that \emph{whatever the
  correct relaxation mechanism at low temperatures, this mechanism is
  at work here,} since on average $80\%$ of the particles have
rearranged at a temperature well below $T_\mathrm{MCT}$.  Moreover,
$q(R)$ does not seem to decay exponentially to $\epsilon$ for large
$R$, but rather as a power law (see later). Finally, there is no clear
indication of a plateu corresponding to the self-overlap
$q_{\alpha\alpha}$.  Unsurprisingly, a mosaic fit of the data, using
Eqs.~\ref{jogabonito} and~\ref{barnetta}, gives unphysical values of
the paramenters, in particular a negative value for the complexity
$\Sigma$ and of the surface tension $\Upsilon$.

\begin{figure}
\includegraphics[clip,width=.75\columnwidth,angle=270]{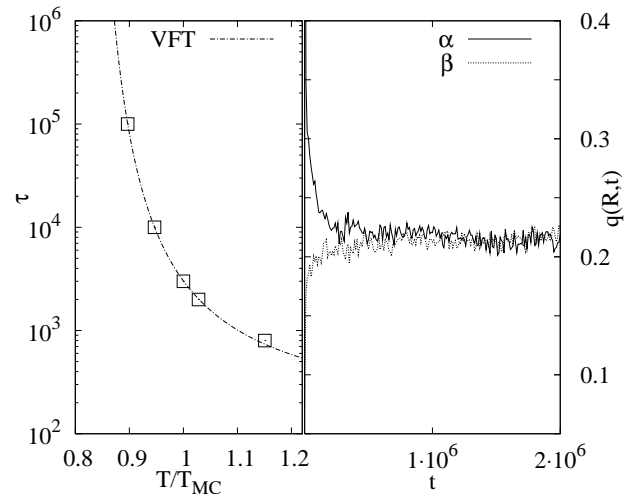}
\caption{Left: Relaxation time {\sl vs.\/}\ temperature for MC swap
  dynamics and VFT fit. $\tau$ is an integrated relaxation time
  obtained from the coarse-grained density autocorrelation studied in
  ref.~\onlinecite{Grigera04} .  Right: Instantaneous overlap {\sl
  vs.\/}\ time for $\alpha$ and $\beta$ initial configurations of the
  sphere ($T=0.89\tmc$, $M=5500$). }
\label{isteresi}
\end{figure}

Three objections can be raised at this point. First, the spheres are
not large enough, the mosaic drop takes place for $\ximos$ larger than
our largest $R$. To this there are two replies: first, $\ximos$ should
{\it decrease} for increasing $T$, so it is unclear why we do not see
anything while approaching $\tmc$ from below; second: as already said,
the largest sphere has largely decorrelated, so, if the mosaic bit of
decorrelation still has to come, it will contribute only to $20\%$ of
the rearrangement.  Second objection: the equilibration time of the
sphere is longer than our runs, and our results for $q(R)$ are
obtained in a {\em metastable} region.  To check this we have repeated
the semulations but initializing the sphere in a different equilibrium
state $\beta$, having zero overlap with the reference state $\alpha$.
What we find (Fig.~\ref{isteresi}, right) is that the overlap as a
function of time {\it increases}, thermalizing at the same value as
when starting within the same state $\alpha$. Thus, even for the
largest $R$, the asymptotic value of $q(R)$ does not depend on the
initial configuration of the sphere, and there is no hysteresis.
Third objection: the lowest temperature is not low enough, it is too
close to $\tmc$, and the mosaic mechanism is not yet at work. We
cannot exclude this. However, at our lowest temperature standard
molecular dynamics is completely stuck, so one would expect the
\ac{MS} to describe the relaxation. Besides, an estimate of $T_0$ from
a VFT fit of the relaxation time gives $T_0 \approx 0.80\tmc$ (see
Fig.~\ref{isteresi}, left), so one would really expect the mosaic
mechanism to be operating at these temperatures.

\begin{figure}
\includegraphics[clip,width=.75\columnwidth,angle=270]{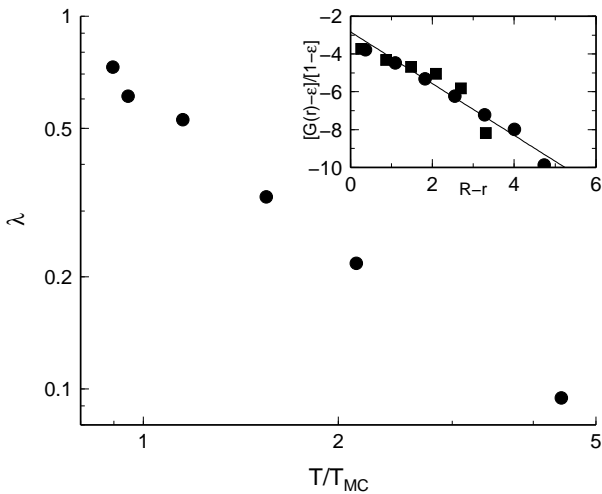}
\caption{Correlation length $\lambda$ vs. $T$. Inset:
  $[G(r)-\epsilon]/(1-\epsilon)$ vs. $R-r$ for $T=0.89\tmc$ at $M=400$
  and $M=5500$.  The line is our prediction, Eq.~\ref{karlo}.}
\label{Gr}
\end{figure}

If we give up the key assumption of the \ac{MS}, that is the existence
of many states at low temperature, we can interpret our data as
relaxation within a single state with self-overlap equal to
$\epsilon$, i.e.  the liquid state. The argument goes as follows. Let
us divide the sphere in shells of radius $r$. From Eq.~\ref{elizondo}
we have,
\begin{equation}
  q(R) = \frac{1}{\frac{4}{3} \pi R^3
  \rho}\int _0^R \!\! dr \, \frac{4 \pi r^2}{l^3} \langle q(r)\rangle =
   \frac{3}{R^3}\int _0^R \!\!dr \, r^2  G(r),
\label{boskov}
\end{equation}
where $G(r)=\langle q(r)\rangle/\epsilon$ is the average overlap per
unit volume between the state $\alpha$ and the asymptotic pinned
state, at distance $r$ from the centre of the sphere.  The effect of
the pinning border at the centre of the sphere is expected to decay as
$\exp(-R/\lambda)$, where $\lambda$ is a correlation
length~\cite{parisi}.  If we say that the unpinned state (the liquid)
has self-overlap $\epsilon$, and we assume $G(r=R)=1$ (sticky
boundary), a reasonable form for $G(r)$ is given by,
\begin{equation}
      G(r) = (1-\epsilon) \; e^{-(R-r)/\lambda} + \epsilon.
\label{karlo}
\end{equation}
Plugging $G(r)$ into Eq.~\ref{boskov}, and having defined $x\equiv
R/\lambda$, we obtain the one-state (1S) overlap,
\begin{equation}
  q_\mathrm{1S}(R) = 3(1-\epsilon)\left[ \frac{1}{x} - \frac{2}{x^2} + \frac{2
      \left(1- e^{-x}\right)}{x^3} \right] + \epsilon .
\label{naive}
\end{equation}
This function describes the relaxation within a single state with
self-overlap $\epsilon$, and it is quite different from the \ac{MS}
form (Eq.~\ref{barnetta}). In particular, $q_\mathrm{1S}(R) - \epsilon
\sim \lambda/R$, for large $R$. In Fig.~\ref{qR} we report the fit to
the data obtained with the 1S overlap, which seems quite reasonable
\footnote{The argument breaks down at low values of $R$, where there
  is no difference between surface and bulk.}.  The only fitting
parameter is the correlation length $\lambda$, which increases by a
factor $7$ with decreasing $T$ in our temperature span
(Fig.~\ref{Gr}).

We can test assumption~(\ref{karlo}) by studying, at fixed $R$, the
average overlap as a function of the distance $r$ from the center
(Fig.~\ref{Gr}, inset). We plot $\log[(G(r)-\epsilon)/(1-\epsilon)]$
vs.  $R-r$, for two different values of $R$, at our lowest $T$.  The
data are compared to our guess Eq.~\ref{karlo}, with $\lambda$
obtained from the fit of $q(R)$. The agreement is reasonable. Another
test is to plot $[q_\mathrm{1S}(R)-\epsilon] / (3-3\epsilon)$ vs.\
$R/\lambda$, where according to~(\ref{naive}) data for all
temperatures should fall on a single master curve. Again
(Fig.~\ref{qR}, inset) we find a nice agreement. We note that the
correlation function $G(r)$ is a kind of the ``point-to-set''
correlation functions described in~\cite{MontanariSem06}.  Other than
testing the \ac{MS}, it seems from our results that $G(r)$ may be a
useful tool to detect growing spatial correlations in supercooled
liquids.

The one-state decay of $q(R)$ described above should in fact be used
also in the small $R$ regime of the mosaic scenario. This modifies
Eq.~\ref{barnetta} as follows:
\begin{equation}
  q_\mathrm{ms}(R) =
  p_{\alpha\alpha}(R)\, q_{1\mathrm{S}}(R) + p_{\alpha\beta}(R) \epsilon
\label{barnetta2}
\end{equation}
where now the limiting self-overlap of $q_{1\mathrm{S}}(R)$ for
$R\to\infty$ is $q_{\alpha\alpha}$, hopefully larger than $\epsilon$
in a multi-state scenario.  Thus, the correct mosaic prediction is
that the relaxation of the sphere within state $\alpha$ is ruled for
small $R$ by $q_{1\mathrm{S}}(R)$, but it is interrupted at $R=\ximos$
by the entropic drive kicking in, such that $q(R)\to \epsilon$ for $R
>\xi_\mathrm{ms}$. We have tried Eq.~\ref{barnetta2} to fit the data
of Fig.~\ref{qR}, leaving $\theta$ fixed to avoid having too many
parameters. For $\theta=2$ and $\theta=3/2$ (two values one might
expect \cite{Xia01}), the fits locate the mosaic drop just beyond the
largest simulated size, with $q_{\alpha\alpha}\sim\epsilon$.  This
makes the mosaic fit rather awkward. In our opinion, given the present
data, it is more natural to conclude that $\alpha$ is the (unique)
liquid state, with self-overlap $\epsilon$, and that there is no
mosaic drop beyond our largest $R$.

We have studied the themodynamics of a sphere embedded in a frozen
equilibrium environment. Our data are compatible with the naive
expectation that the inner shells of the sphere decorrelate and
rearrange more than the outer shells close to the border. The
penetration scale of the pinning border is given by a length
$\lambda$, which increases sharply with decreasing temperature. In the
paramagnetic phase of the Ising model, $\lambda$ diverges for $T\to
T_c$, just as the standard correlation length~\cite{Cammarota06}. This
means that $\lambda$ has physical significance and begs for an
investigation of the relation between $\lambda$ and $\tau$ also in
glass-formers.  Such an investigation, however, would require a finer
resolution of $\lambda(T)$ around and below $\tmc$, which we leave for
a future study.  The length $\lambda$ is neatly defined through the
novel correlation function $G(r)$. Our study is purely thermodynamic,
and for this reason we do not see any crossover at the \acl{TMC}.

Our data do not show any clear evidence for the mosaic mechanism.
There is the possibility that the experiment or the observable
considered are not the most suitable. For example, if the mosaic
excitations are highly non-compact or fractal~\cite{Stevenson06}, it
may be argued that a different experiment would be needed. Having said
that, we remain with the impression that the one-state scenario fits
reasonably well our data.  Either there is no surface tension, or it
is very weak and diluted over a wide region across the border. This
suggests that, if a multi-state scenario is valid, a version more
sophisticated than the here tested is needed.  This novel description
should not only reproduce the numerical data, but fit them
significantly better than the single-state scenario.

We thank G. Biroli, J.-P. Bouchaud, S. Franz and F. Zamponi for
discussions and I. Giardina for important suggestions.  TSG thanks
SMC-INFM-CNR (Rome) for hospitality and Fundaci\'on Antorchas, ANPCyT,
CONICET, and UNLP (Argentina) for financial support.

\bibliography{biblio}

\acrodef{MCT}{mode coupling theory}
\acrodef{MS}[MS]{mosaic scenario}
\acrodef{TMC}[$\tmc$]{mode coupling temperature}

\end{document}